\documentclass[graybox]{svmult}
\usepackage[english]{babel}
\usepackage[utf8x]{inputenc}
\usepackage[T1]{fontenc}
\usepackage{amsmath}
\usepackage{graphicx}
\usepackage[usenames,dvipsnames]{xcolor}
\usepackage[colorinlistoftodos]{todonotes}
\usepackage{listings}
\usepackage{todonotes}

\usepackage{mathptmx}      
\usepackage{helvet}        
\usepackage{courier}        
\usepackage{type1cm}                               
\usepackage{makeidx}        
\usepackage{graphicx}                                     
\usepackage{multicol}        
\usepackage[bottom]{footmisc}

\definecolor{color:keyword}{rgb}{0.53,0.05,0.05}
\definecolor{color:comment}{rgb}{0.25,0.37,0.75}
\definecolor{color:string}{rgb}{0.87,0.0,0.0}
\usepackage{bold-extra}

\lstdefinelanguage{Jolie}{
morekeywords={
	provide,until,OneWay,RequestResponse,new,
	main,define,inputPort,outputPort,init,execution,include,
	cset,if,else,csets,interface,type,throws,global,constants,for,
foreach,while,int,double,raw,void,undefined,string,long,bool,any,single,
sequential,concurrent,Jolie,Java,JavaScript,embedded,Location,Protocol,
Interfaces,Aggregates,scope,install,cH,comp,throw,this,default,synchronized,
nullProcess,false,true
},
sensitive=true,
morecomment=[l]{//},
morecomment=[s]{/*}{*/},
morestring=[b]",
otherkeywords={;,|,:}
}

\makeindex  

\begin{document}

\title*{Size Matters: Microservices Research and Applications}
\author{Manuel Mazzara, Antonio Bucchiarone, Nicola Dragoni, Victor	Rivera}
\institute{Antonio Bucchiarone \at Fondazione Bruno Kessler, Trento, Italy \email{bucchiarone@fbk.eu}
\and  Manuel Mazzara, Victor Rivera \at Innopolis University, Russian Federation \email{{m.mazzara, v.rivera}@innopolis.ru}
\and  Nicola Dragoni \at DTU Compute, Technical University of Denmark and Centre for Applied Autonomous Sensor Systems
Orebro University, Sweden \email{ndra@dtu.dk}
}

\authorrunning{Mazzara et al.}

\maketitle

 \abstract{
In this chapter we offer an overview of microservices providing the introductory information that a reader should know before continuing reading this book. We introduce the idea of microservices and we discuss some of the current research challenges and real-life software applications where the microservice paradigm play a key role. We have identified a set of areas where both researcher and developer can propose new ideas and technical solutions.}

\section{The Shift Towards Distribution}

History of programming languages, paradigms and software architectures have been characterized in the last few decades by a progressive shift towards distribution, modularization and loose coupling. The purpose is to increase code reuse and robustness \cite{AlmeidaALGM04,JamshidiPMLT18}, ultimately a necessity dictated by the need of increasing software quality, not only in safety and financial-critical applications \cite{MazzaraTSC2018}, but also in more common off-the-shelf software packages. The two directions of \textit{modularization} \footnote{\url{https://www.oreilly.com/ideas/modules-vs-microservices}} (code reuse and solid design) and \textit{robustness} (software quality and formal methods: verification/correctness-by-construction) advanced to some extent independently and pushed by different communities, although with a non-empty overlap.

Object-oriented technologies are prominent in software development \cite{rumbaugh1991object}, with specific instances of languages incorporating both the aspects aforementioned (modularity and correctness). A notable example is the Eiffel programming language \cite{Meyer1988}, incorporating solid principles of Object-Oriented-Programming (OOP) within a programming framework coordinated by the idea of \textit{design-by-contract}, which aims at correctness-by-construction. None of these technologies can nevertheless rule out the need for testing, which robustly remains a pillar of the software development lifecycle.  

Other examples exist of languages having a strong emphasis on correctness, both from the architectural viewpoint and in terms of meeting functional requirements \cite{Mazzara2010}. However, until recently, not much attention was dedicated to integrating these principles into a distributed setting winning out properties such as easiness of deployment, a lightweight design and development phase, and minimal need for integration testing. The idea of Microservices \cite{Dragoni2017,JamshidiPMLT18} and Devops \cite{BassDevOpsA2015,Bobrov19,MazzaraNSSU18}  stem out exactly from this widespread and recognized need.

\emph{Chapter Outline and Contribution.} The contribution of the chapter is twofold, and thus organized in two main sections. Section~\ref{sec:Microservices} overviews the essential concepts characterizing the Microservices paradigm, thus serving as introduction for the entire book. Section~\ref{sec:Research} instead highlights some key research areas in which Microservices applications have gained particular interest and showed some research progress. Conclusions and future works are summed up in Section \ref{sec:Conclusions}.

\section{Microservices}
\label{sec:Microservices}

Microservices~\cite{Dragoni2017,F14,JamshidiPMLT18,N15} is an architectural style stemming from Service-Oriented Architectures (SOAs)~\cite{mackenzie2006,sillitti2002}. According to this architectural style, a system is structured by  small independent building blocks -- the \textit{microservices} -- communicating only via message passing. The main idea is to move \textit{in the small} (within an application) some of the concepts that worked \textit{in the large}, i.e. for cross-organization business-to-business workflow which makes use of orchestration engines such as WS-BPEL (in turn inheriting some of the functional principles from concurrency theory~\cite{LucchiM07}). The characteristic differentiating the new style from monolithic architectures and classic service-oriented architectures is the emphasis on \textit{scalability}, \textit{independence}, and \textit{semantic cohesiveness} of each unit constituting the system. In its fundamental essence, the Microservices architecture \cite{DragoniGLMMMS17} is built on a few very simple principles:

\begin{itemize}
	\item \textit{Bounded Context}. First introduced in \cite{evans2004domain}, this concept captures one of the key properties of Microservices architecture: focus on business capabilities. Related functionalities are combined into a single business capability which is then implemented as a service.
	\item \textit{Size}. Size is a crucial concept for microservices and brings major benefits in terms of service maintainability and extendability. Idiomatic use of microservices architecture suggests that if a service is too large, it should be refined into two or more services, thus preserving granularity and maintaining focus on providing only a single business capability.
	\item \textit{Independency}. This concept encourages loose coupling and high cohesion by stating that each service in Microservices architectures is operationally independent from others, and the only form of communication between services is through their published interfaces. 
\end{itemize}

\subsection{Microservices vs. Monolith}

\begin{figure}[!ht]
\centering
\includegraphics[width=0.4\textwidth]{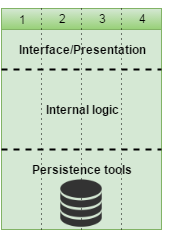}
\caption{Monolith Architecture \label{comparison}}
\end{figure}

All the programming languages for development of server-side applications provide abstractions to break down the complexity of programs into modules or components \cite{clark2004,gross2005,predonzani2001}. However, these languages are designed for the creation of single executable artifacts. In monolithic architectures, the modularization abstractions rely on the sharing of resources such as memory, databases and files of the same machine. The components are therefore not independently executable. Figure \ref{comparison} (reproduced from \cite{MazzaraSSE}) shows the classic monolithic organization: the different layers of the system (interface/presentation, internal business logic, and persistence tools) are here split in terms of responsibilities between different modules (the vertical split with numbers from 1 to 4). In fact, each module may take part in the implementation of functionalities related to each layer, the database is common, and so the access to other resources such as memory.  

\begin{figure}[!ht]
\centering
\includegraphics[width=0.75\textwidth]{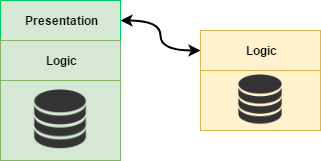}
\caption{Microservices Architecture \label{ms}}
\end{figure}

Figure \ref{ms} (reproduced from \cite{MazzaraSSE}) shows the componentization in a Microservices architecture. Each service has its own dedicated persistence tool and communication is via message passing. With this organization there is no vertical split through all the system layers, and the deployment is independent. The complexity is moved to the level of coordination of services (often called orchestration~\cite{Mazzara2005}). Moreover, a number of additional problems need to be addressed due to the distributed nature of the Microservices approach (e.g., trust and certification \cite{damiani2009,DRAGONI20091628}).

\subsection{Microservices vs. SOA}

In SOA, services are not required to be self-contained with data and User Interface, and their own persistence tools, eg. database. SOA has no focus on independent deployment units and related consequences, it is simply an approach for business-to-business intercommunication. The idea of SOA was to enable business-level programming through business processing engines and languages such as WS-BPEL and BPMN that were built on top of the vast literature on business modelling~\cite{YanMCU07}. Furthermore, the emphasis was all on \textit{service composition} \cite{LemosDB16,DustdarS05} more than service development and deployment.

\subsection{Size Matters: The Organization of Teams}

A microservice is not just a \textit{very small service}. There is no predefined size limit that defines whether a service is a microservice or not. From this angle, the term ``microservice'' can somehow be misleading. Each microservice is expected to implement a single \textit{business capability}, in fact a very limited system functionality, bringing benefits in terms of service maintainability and extendability. Since each microservice represents a single business capability, which is delivered and updated independently, discovering  bugs or adding minor improvements do not have any impact on other services and on their releases. In common practice, it is also expected that a single service can be developed and managed by a single team~\cite{Dragoni2017}. 

In order to build a system with a modular and loosely coupled design, it is necessary to pay attention to the organization structure and the communication patterns. These patterns directly impact the produced design (Conway's Law~\cite{conway1968committees}. If a structure is based on the idea that each team work on a single service, then the communication will be more efficient at the team level and in the entire organization. This will lead to an improved design in terms of modularity. Microservices' approach is to keep teams small and communications efficient by creating small cross-functional (DevOps) teams that are able to continuously work on the same service and to be fully responsible for it (``you build it, you run it'' principle~\cite{gray2006conversation}). 

The teams are organized around services, which in turn are organized around business capabilities~\cite{F14}. The optimal team size for microservices is best described by Jeff Bezos’ famous ``two pizza team'' rule, which suggests that the size of a team should be no larger than what two pizzas can feed. The rule itself does not give an exact number, however it is possible to estimate it to be around 6-8 people. The drawback of such approach is that it is not always practical from the financial point of view to maintain a dedicated team of developers for a single service as it may lead to high development/maintenance costs~\cite{jones2014soa}. Furthermore, one should be careful when designing the high level structure of the organization using microservices - increasing  the number of services might negatively impact the overall organization efficiency, if no further actions are taken. 

\section{Research and Applications}
\label{sec:Research}

Microservices have recently seen a dramatic growth in popularity and in concrete applications~\cite{N15}. The shift towards microservices is seeing several companies involved in a major refactoring of their back-end systems to accommodate the new paradigm \cite{BucchiaroneDDLM18, MazzaraTSC2018}. Other companies just start their business model developing software following the microservice paradigm since day one. We are in the middle of a major change in the view in which software is intended, and in the way in which capabilities are organized into components, and industrial systems are conceived. In this section we describe recent research progress for what concern Microservices applications \cite{MazzaraSSE}. It can be structured in the following areas:

\begin{itemize}
	\item \textit{Programming Languages}
	\item \textit{Type Checker}
	\item \textit{Migration from Monoliths}
	\item \textit{Education in DevOps}
	\item \textit{Modeling and Self-adaptability}
    \item \textit{Real-life software applications with Microservices}
\end{itemize}

\subsection{Programming Languages}

Microservice systems are currently developed using mostly general-purpose programming
languages that do not provide dedicated abstractions for service composition. Current practice is indeed focused on the deployment aspects of microservices, in particular by using containerization. We investigated this issue and made a case for a language-based approach to the engineering of Microservices architectures. We believe that this approach is complementary to current practice. In \cite{Guidi2017} we discussed the approach in general, and we instantiate it in terms of the Jolie programming language; however the concept is independent from the specific technical solution adopted. Four important concepts have been identified to be first class entities in the programming language in order to address the Microservices architecture:

\begin{enumerate}
\item \textit{Interfaces}: to support modular programming, services has to be deployed as \textit{black boxes}. In order to compose services in larger systems, interfaces have to describe the provided functionalities and those required from the environment.
\item \textit{Ports}: since a microservice interacts with other services, a communication port describes how its functionalities are made available to the network (interface, communication technology, and data protocol). Ports should be specified separately from the implementation of a service. Input ports describe the functionalities that the service provides to the rest of the system, while output ports describe the functionalities that the service requires from the rest of the system.
\item \textit{Workflows}: structured protocols appear repeatedly in microservices and they are not natively supported by mainstream languages. All possible operations are always enabled (for example in Object-Oriented programming). Causal dependencies are programmed by using a book-keeping variable, which is error-prone, and it does not scale when the number of causality links increases. A microservice language should provide abstractions for programming workflows.
\item \textit{Processes}: workflows define the blueprint of the behavior of a service. At runtime a service may interact with multiple clients and other external services, therefore there is need to support multiple concurrent executions of its workflow. A process is a running instance of a workflow, and a service may include many processes executing concurrently. Each process runs independently of the others, to avoid interference, and has its own private state.
\end{enumerate}

\subsection{Type Checker}

Static type checking is generally desirable for programming languages improving software quality, lowering the number of bugs and preventing avoidable errors. The idea is to allow compilers to identify as many issues as possible before actually run the program, and therefore avoid a vast number of trivial bugs, catching them at a very early stage. Despite the fact that, in the general case interesting properties of programs are undecidable \cite{Rice53}, static type checking, within its limits, is an effective and well established technique of program verification. If a compiler can prove that a program is well-typed, then it does not need to perform dynamic safety checks, allowing the resulting compiled binary to run faster. 

In \cite{JSTC} we described and prototyped the Jolie Static Type Checker (JSTC), a static type checker for the Jolie programming language which natively supports microservices. The static type system for the language was exhaustively and formally defined on paper \cite{nielsen}, but needed implementation. The type checker prototype consists of a set of rules for the type system expressed in SMT Lib language. The actual implementation covers operations such as assignments, logical statements, conditions, literals and comparisons.

In \cite{reftype} we integrated dynamic and static type checking with the introduction of refinement types, verified via an SMT solver. The integration of the two aspects allows a scenario where the static verification of internal services and the dynamic verification of (potentially malicious) external services cooperate in order to reduce testing effort and enhance security.

\subsection{Migration from Monoliths}

Several companies are evaluating pros and cons of a migrating to microservices. Financial institutions are positioned in a difficult situation due to the economic climate and the appearance of small players that grew big fast in recent times, such as alternative payment systems, that can also navigate in a more flexible (and less regulated) legal framework and started their business since day one with more agile architectures and without being bounded to outdated technological standard. We worked closely with Danske Bank, the largest bank in Denmark and one of the leading financial institutions in Northern Europe, to demonstrate how scalability is positively affected by re-implementing a monolithic architecture into a Microservices one \cite{BucchiaroneDDLM18}. 

Evolution is necessary to stay competitive. When compared with companies (such as Paypal) that started their activities using innovative technologies as a business foundation, in order to scale and deliver value, old banking institutions appear outdated with regards to technology standards. We worked on the \textit{FX Core} system, a mission critical system of Danske Bank's software. A key outcome of our research has been the identification of a repeatable migration process that can be used to convert a real world Monolithic architecture into a Microservices one in the specific setting of a financial system, which is typically characterized by legacy systems and batch-based processing on heterogeneous data sources~\cite{MazzaraTSC2018}.

\subsection{Education in DevOps}

DevOps is a natural evolution of the Agile approaches \cite{DevOpsHandbook,Bass} from the software itself to the overall infrastructure and operations. This evolution was made possible by the spread of cloud-based technologies and the everything-as-a-service approaches. Adopting the DevOps paradigm helps software teams to release applications faster and with more quality. DevOps and Microservice Architecture appear to be an indivisible pair for organizations aiming at delivering applications and services at high velocity. Investing in DevOps is a good idea in general, and after a migration to microservices it is typically crucial.

As long as DevOps became a widespread philosophy, the necessity of education in the field become more and more important, both from the technical and organisational point of view \cite{BucenaK17}. The DevOps philosophy may be introduced in companies with adequate training, but only if certain technological, organizational and cultural prerequisites are present. If not, the prerequisites should be developed. We have been deeply involved in recent years in teaching both undergraduate and graduate students at the university, and junior/senior professional developers in industry. We have been also working often with management \cite{MazzaraNSSU18,Bobrov19}.

\subsection{Modeling and Self-Adaptability}

Innovative engineering is always looking for adequate tools to model and verify software systems, as well as support developers in deploying correct software. Microservices is an effective paradigm to cope with scalability; however, the paradigm still misses a \textit{conceptual model} able to support engineers since the early phases of development. To make the engineering process of a microservices-based application efficient, we need a \textit{uniform way to model autonomous and heterogeneous microservices}, at a level of abstraction that allows for easy interconnection through dynamic relations. Each microservice must have a partial view on the surrounding operational environment (i.e., system knowledge) and at the same time must be able to be specialized/refined and adapted to face different requirements, user needs, context-changes, and missing functionalities.

To be robust, each microservice must be able to dynamically adapt its behaviour and its goals to changes in the environment but also to collaborative interactions with other microservice during their composition/orchestration. At the same time the adaptation must not be controlled centrally and imposed by the system but must be administrated in a decentralised fashion among the microservices.

An important feature of dynamic and context-aware service-based systems is the possibility of handling at run-time extraordinary/improbable situations (e.g., context changes, availability of functionalities, trust negotiation), instead of analyzing such situations at design-time and pre-embedding the corresponding recovery activities. The intrinsic characteristics of microservice architectures make possible to nicely model run-time dependability concepts, such as ``self-protecting'' and ``self-healing'' systems \cite{DRAGONI20091628}. To make this feasible, we should enable microservices to monitor their operational environment and trigger adaptation needs each time a specific system property is violated. To cover the aforementioned research challenges, we already started to define a roadmap \cite{roadmap} that includes an initial investigation on how Domain Objects \cite{BucchiaroneSMPT16} could be an adequate formalism both to capture the peculiarity of MSA, and to support the software development since the early stages.

\subsection{Real-life Software Applications with Microservices}

\subsubsection{Smart Buildings}
Smart buildings represent a key example of application domain where properties like scalability, minimality and cohesiveness play a key role. As a result, smart buildings are an ideal application scenario for the Microservices paradigm. This domain has been investigated with an outlook on Internet-of-Things technologies (IoT) and smart cities \cite{MazzaraSDB2019}. In \cite{Salikhov2016b} and \cite{Salikhov2016a} it has been shown how rooms of a building can be equipped with devices and sensors in order to capture the fundamental parameters determining well-being and livability of humans, such as temperature, humidity, and illumination. This solution allows to monitor an equipped area and therefore collect data that can be mined and analyzed for specific purposes. The nodes used in this system consist of Raspberry Pi micro-computers~\cite {Raspberri}, Texas Instruments Sensor Tags~\cite{TIST}, door sensor and a web camera.  Currently, this system is able to collect and analyze room temperature, pressure and illumination level. It is also able to distinguish and count people, which are located in the covered area. The purpose is to monitor and optimize working conditions. The software infrastructure, tightly connected to the hardware, makes use of Microservices to achieve the desired level of scalability, minimality and cohesiveness. Sensors and actuators are connected to a central control panel that is responsible to manage them. At the same time, an automatic Personal Assistant has been designed. It is capable to observe data, learn about different users preferences, and adapt the room conditions accordingly for the different phases of his/her work \cite{KhusnutdinovUMK18}.

\subsubsection{Smart Mobility}
Organizing and managing the mobility services within a city, meeting traveler’s expectations and properly exploiting the available transport resources, is becoming a more and more complex task. The inadequacy of traditional transportation models is proven by the prolification of alternative, social and grassroots initiatives aiming at a more flexible, customized and collective way of organizing transport (e.g., carpooling, ride and park sharing services, flexi-buses) \cite{DakroubBLAA13,Fagnant2016,Furuhata2013}. Some of these attempts have been very successful (e.g., Uber), even if in most cases these are seen as isolated solutions targeting specific mobility target groups and are not part of the city mobility ecosystem, mainly based on traditional public and private transport facilities. 

An attempt of re-thinking the way mobility is managed and offered is represented by the Mobility as a Service (MaaS) model. MaaS solutions (e.g., MaaS Global: http://maas.global) aim at arranging the most suitable transport solution for their customers thanks to cost effective integrated offer of different multi-modal means of transportation. MaaS also foresees radical changes in the business landscape, with a new generation of mobility operators emerging as key actors to manage the increased flexibility and dynamism offered by this new concept of mobility.

People need to travel quickly and conveniently between locations at different scales, ranging from a trip of a few blocks to a journey across town or further. Each trip has its set of requirements. Time may be of the essence. Cost may be paramount, and the convenience of door-to-door travel may be important. In each case, the transportation infrastructure should seamlessy provide the best option. A modern city needs to flexibly integrate transportation options including buses, trains, taxis, autonomous vehicles, bicycles and private cars. 

Before changing communities to support what is believed the future transportation will look like and behave, it is necessary to develop mechanisms that allow planners of these localities to model, analyse, and present these possible configurations in ways that the citizens of the communities can understand and participate in.

\textit{Coordination for Mobility as a Service} can be implemented on a spectrum, ranging from independent services communicating exclusively through market exchanges to hybrid market/hierarchy approaches fixed hierarchical control systems.

Every transportation mean does not need to be a individual competing across multiple markets, but neither should there only be one rigid hierarchy. \textit{"Diversity"} and \textit{“Distributed”} selection of the appropriate mean (or a combination of them) is the appropriate compromise respect to say that if one is better than the other, we "kill" the other.

To realize such a \textit{"dynamic"} and \textit{"emergent"} behaviors in transportation systems, needs a new way for developing their supporting software systems. In the last years, Collective adaptive systems (CAS) have been introduced and studied by many researchers in different application domains (i.e., Industry 4.0, Logistics, Smart Cities and Mobility, Energy, Biology, etc..) \footnote{\url{http://www.focas.eu/focas-manifesto.pdf}}.
CAS consists of diverse heterogeneous entities composing a socio-technical system. Individual entities "opportunistically" enter a system and self-adapt in order to leverage other entities’ resources and services to perform their task more efficiently or effectively. At the same time, also collections of entities, called Ensembles, must be able to self-adapt simultaneously to preserve the collaboration and benefits of the system (or sub-system) they are within.

In this very dynamic and rapidly evolving setting, microservices have the potential of offering the right concepts for modeling and for programming smart mobility solutions. Coordination for Mobility as a Services (MaaS) is a mandatory requirement to maintain a certain level of city sustainability (i.e., less CO2 emission, more citizen participation and satisfaction, etc..).  It can be implemented on a spectrum, ranging from independent agents communicating exclusively through market exchanges to hybrid market/hierarchy approaches fixed hierarchical control systems. Our opinion is that instead to implement a selfish mobility we see the need to realize a collective and cooperative mobility where each MaaS provider sees in each single competitors a partner and not an enemy \cite{BucchiaroneSM16}. This domain open new challenges in how distributed microservices, provided by different mobility entities, can be composed dynamically to provide real-time and continuous answers to citizens in a Smart City.

\section{Conclusions}
\label{sec:Conclusions}

The microservice architecture is a style that is increasingly gaining popularity, both in academia and in the industry. Even though it is likely to conduct to a paradigm shift and a dramatic change in perception, it does not build on vacuum, and instead relates to well-established paradigms such as OO and SOA. In \cite{DragoniGLMMMS17} a comprehensive survey on recent developments of Microservices architecture is presented with focus on the \emph{evolutionary} aspects more than the \emph{revolutionary} ones. The presentation there is intended to help the reader in understanding the distinguishing characteristics of microservices.

We have a long experience in the field of services and business processes \cite{BucchiaroneMPR17, LaneBR12, mazzaraPhD, YanCZM07, YanMCU07}, including workflows and their reconfiguration \cite{BucchiaroneMPR12,MDZ2011, Mazzara11}. We built on top of this expertise to focus on the active research field of Microservices, and summarized our work in this chapter.

The future will see a growing attention regarding the matters discussed in this chapter, and the development of new programming languages intended to address the microservice paradigm~\cite{Guidi2017}. Object-Oriented programming brought fresh ideas in the last decades, and the expectation is that a comparable shift may be just ahead of us. Holding on the optimism the future is certainly not challenge-free. Security of the microservice paradigm is an issue almost fully untouched \cite{DragoniGLMMMS17}. Commercial-level quality packages for development are still far to come, despite the acceleration in the interest regarding the matter. Fully-verified software is an open problem the same way it is for more traditional development models. That said, several research centers around the world have addressed and are addressing all these issues in the attempt to ride the wave and make the new generation of distributed systems a reality.

\bibliographystyle{plain}
\bibliography{MS}  

\begin{thebibliography}{10}

\bibitem{Raspberri}
Raspberri pi official site.
\newblock \url{https://www.raspberrypi.org/}, Last accessed June 2017.

\bibitem{TIST}
Texas instruments sensor tag official site.
\newblock
  \url{http://www.ti.com/ww/en/wireless_connectivity/sensortag/gettingStarted.html},
  Last accessed June 2017.

\bibitem{BassDevOpsA2015}
Len Bass, Ingo Weber, and Liming Zhu.
\newblock {\em DevOps: A Software Architect's Perspective}.
\newblock Addison-Wesley Professional, 2015.

\bibitem{Bass}
Len Bass, Ingo Weber, and Liming Zhu.
\newblock {\em DevOps: A Software Architect's Perspective}.
\newblock Addison-Wesley Professional, 1st edition, 2015.

\bibitem{Bobrov19}
Evgeny Bobrov, Antonio Bucchiarone, Alfredo Capozucca, Nicolas Guelfi, Manuel
  Mazzara, and Sergey Masyagin.
\newblock Teaching devops in academia and industry: reflections and vision.
\newblock In {\em Software Engineering Aspects of Continuous Development and
  New Paradigms of Software Production and Deployment - Second International
  Workshop, {DEVOPS} 2019, Chateau de Villebrumier,}, 2019.

\bibitem{BucchiaroneDDLM18}
Antonio Bucchiarone, Nicola Dragoni, Schahram Dustdar, Stephan~Thordal Larsen,
  and Manuel Mazzara.
\newblock From monolithic to microservices: An experience report from the
  banking domain.
\newblock {\em {IEEE} Software}, 35(3):50--55, 2018.

\bibitem{BucchiaroneMPR12}
Antonio Bucchiarone, Annapaola Marconi, Marco Pistore, and Heorhi Raik.
\newblock Dynamic adaptation of fragment-based and context-aware business
  processes.
\newblock In {\em {ICWS}}, pages 33--41. {IEEE} Computer Society, 2012.

\bibitem{BucchiaroneMPR17}
Antonio Bucchiarone, Annapaola Marconi, Marco Pistore, and Heorhi Raik.
\newblock A context-aware framework for dynamic composition of process
  fragments in the internet of services.
\newblock {\em J. Internet Services and Applications}, 8(1):6:1--6:23, 2017.

\bibitem{BucchiaroneSM16}
Antonio Bucchiarone, Martina~De Sanctis, and Annapaola Marconi.
\newblock Decentralized dynamic adaptation for service-based collective
  adaptive systems.
\newblock In {\em {ICSOC} Workshops}, volume 10380 of {\em Lecture Notes in
  Computer Science}, pages 5--20. Springer, 2016.

\bibitem{BucchiaroneSMPT16}
Antonio Bucchiarone, Martina~De Sanctis, Annapaola Marconi, Marco Pistore, and
  Paolo Traverso.
\newblock Incremental composition for adaptive by-design service based systems.
\newblock In {\em {IEEE} {ICWS} 2016, San Francisco, CA, USA, June 27 - July
  2}, pages 236--243, 2016.

\bibitem{BucenaK17}
Ineta Bucena and Marite Kirikova.
\newblock Simplifying the devops adoption process.
\newblock In {\em Joint Proceedings of the {BIR} 2017 pre-BIR Forum, Workshops
  and Doctoral Consortium co-located with 16th International Conference on
  Perspectives in Business Informatics Research {(BIR} 2017), Copenhagen,
  Denmark, August 28 - 30, 2017.}, 2017.

\bibitem{clark2004}
{Clark J.}, {Clarke C.}, {De Panfilis S.}, {De Panfilis S.}, {Sillitti A.},
  {Succi G.}, and {Vernazza T.}
\newblock Selecting components in large {COTS} repositories.
\newblock {\em Journal of Systems and Software}, pages 323 -- 331, 2004.

\bibitem{conway1968committees}
Melvin~E Conway.
\newblock How do committees invent.
\newblock {\em Datamation}, 14(4):28--31, 1968.

\bibitem{DakroubBLAA13}
Oussama Dakroub, Carl~Michael Boukhater, Fayez Lahoud, Mariette Awad, and
  Hassan Artail.
\newblock An intelligent carpooling app for a green social solution to traffic
  and parking congestions.
\newblock In {\em 16th International {IEEE} Conference on Intelligent
  Transportation Systems, {ITSC} 2013, The Hague, The Netherlands, October 6-9,
  2013}, pages 2401--2408, 2013.

\bibitem{damiani2009}
{Damiani E.}, {El Ioini N.}, {Sillitti A.}, and {Succi G.}
\newblock {WS}-certificate.
\newblock In {\em 2009 {IEEE} International Workshop on Web Services Security
  Management}. {IEEE}, 2009.

\bibitem{JSTC}
Bogdan Mingela Larisa Safina Alexander Tchitchigin Nikolay~Troshkov Daniel~de
  Carvalho, Manuel~Mazzara.
\newblock Jolie static type checker: a prototype.
\newblock {\em Modeling and Analysis of Information Systems}, 24(6):704–717,
  2017.

\bibitem{AlmeidaALGM04}
Eduardo~Santana de~Almeida, Alexandre Alvaro, Daniel Lucr{\'{e}}dio,
  Vinicius~Cardoso Garcia, and Silvio~Romero de~Lemos~Meira.
\newblock Rise project: Towards a robust framework for software reuse.
\newblock In {\em Proceedings of the 2004 {IEEE} International Conference on
  Information Reuse and Integration, {IRI} - 2004, November 8-10, 2004, Las
  Vegas Hilton, Las Vegas, NV, {USA}}, pages 48--53, 2004.

\bibitem{Dragoni2017}
N.~Dragoni, S.~Giallorenzo, A.~Lluch-Lafuente, M.~Mazzara, F.~Montesi,
  R.~Mustafin, and L.~Safina.
\newblock Microservices: yesterday, today, and tomorrow.
\newblock In {\em Present and Ulterior Software Engineering}. Springer, 2017.

\bibitem{DragoniGLMMMS17}
Nicola Dragoni, Saverio Giallorenzo, Alberto Lluch{-}Lafuente, Manuel Mazzara,
  Fabrizio Montesi, Ruslan Mustafin, and Larisa Safina.
\newblock Microservices: Yesterday, today, and tomorrow.
\newblock In {\em Present and Ulterior Software Engineering.}, pages 195--216.
  2017.

\bibitem{DRAGONI20091628}
Nicola Dragoni, Fabio Massacci, and Ayda Saidane.
\newblock A self-protecting and self-healing framework for negotiating services
  and trust in autonomic communication systems.
\newblock {\em Computer Networks}, 53(10):1628 -- 1648, 2009.

\bibitem{DustdarS05}
Schahram Dustdar and Wolfgang Schreiner.
\newblock A survey on web services composition.
\newblock {\em {IJWGS}}, 1(1):1--30, 2005.

\bibitem{evans2004domain}
Eric Evans.
\newblock {\em Domain-driven design: tackling complexity in the heart of
  software}.
\newblock Addison-Wesley Professional, 2004.

\bibitem{Fagnant2016}
Daniel Fagnant and Kara Kockelman.
\newblock Dynamic ride-sharing and fleet sizing for a system of shared
  autonomous vehicles in austin, texas.
\newblock {\em Transportation}, 45:28--46, 08 2016.

\bibitem{F14}
M.~Fowler and J.~Lewis.
\newblock Microservices.
\newblock {\em ThoughtWorks}, 2014.

\bibitem{Furuhata2013}
Masabumi Furuhata, Maged Dessouky, Fernando Ordóñez, Marc-Etienne Brunet,
  Xiaoqing Wang, and Sven Koenig.
\newblock Ridesharing: The state-of-the-art and future directions.
\newblock {\em Transportation Research Part B: Methodological}, 57:28–46, 11
  2013.

\bibitem{gray2006conversation}
Jim Gray.
\newblock A conversation with werner vogels.
\newblock {\em ACM Queue}, 4(4):14--22, 2006.

\bibitem{gross2005}
{Gross H.G.}, {Melideo M.}, and {Sillitti A.}
\newblock Self certification and trust in component procurement.
\newblock {\em Journal of Science of Computer Programming}, pages 141 -- 156,
  2005.

\bibitem{Guidi2017}
Claudio Guidi, Ivan Lanese, Manuel Mazzara, and Fabrizio Montesi.
\newblock {\em Microservices: A Language-Based Approach}, pages 217--225.
\newblock Springer International Publishing, Cham, 2017.

\bibitem{JamshidiPMLT18}
Pooyan Jamshidi, Claus Pahl, Nabor~C. Mendon{\c{c}}a, James Lewis, and Stefan
  Tilkov.
\newblock Microservices: The journey so far and challenges ahead.
\newblock {\em {IEEE} Software}, 35(3):24--35, 2018.

\bibitem{jones2014soa}
Steve Jones.
\newblock Microservices is soa, for those who know what soa is.
\newblock
  \url{http://service-architecture.blogspot.co.uk/2014/03/microservices-is-soa-for-those-who-know.html},
  2014.

\bibitem{Salikhov2016a}
K.~{Khanda}, D.~{Salikhov}, K.~{Gusmanov}, M.~{Mazzara}, and N.~{Mavridis}.
\newblock Microservice-based iot for smart buildings.
\newblock In {\em 2017 31st International Conference on Advanced Information
  Networking and Applications Workshops (WAINA)}, pages 302--308, March 2017.

\bibitem{KhusnutdinovUMK18}
Azat Khusnutdinov, Denis Usachev, Manuel Mazzara, Adil Khan, and Ivan
  Panchenko.
\newblock Open source platform digital personal assistant.
\newblock In {\em 32nd International Conference on Advanced Information
  Networking and Applications Workshops, {AINA} 2018 workshops, Krakow, Poland,
  May 16-18, 2018}, pages 45--50, 2018.

\bibitem{DevOpsHandbook}
Gene Kim, Patrick Debois, John Willis, and Jez Humble.
\newblock {\em The DevOps Handbook: How to Create World-Class Agility,
  Reliability, and Security in Technology Organizations}.
\newblock IT Revolution Press, 2016.

\bibitem{LaneBR12}
Stephen Lane, Antonio Bucchiarone, and Ita Richardson.
\newblock Soadapt: {A} process reference model for developing adaptable
  service-based applications.
\newblock {\em Information {\&} Software Technology}, 54(3):299--316, 2012.

\bibitem{LemosDB16}
Angel~Lagares Lemos, Florian Daniel, and Boualem Benatallah.
\newblock Web service composition: {A} survey of techniques and tools.
\newblock {\em {ACM} Comput. Surv.}, 48(3):33:1--33:41, 2016.

\bibitem{LucchiM07}
Roberto Lucchi and Manuel Mazzara.
\newblock A pi-calculus based semantics for {WS-BPEL}.
\newblock {\em J. Log. Algebr. Program.}, 70(1):96--118, 2007.

\bibitem{mackenzie2006}
M.C. MacKenzie et~al.
\newblock Reference model for service oriented architecture 1.0.
\newblock {\em OASIS Standard}, 12, 2006.

\bibitem{MDZ2011}
Dragoni Nicola Zhou~Mu. Mazzara, Manuel.
\newblock {Dependable workflow reconfiguration in WS-BPEL}.
\newblock In {\em Proceedings of the 5th Nordic Workshop on Dependability and
  Security}, 2011.

\bibitem{MazzaraTSC2018}
M.~Mazzara, N.~Dragoni, A.~Bucchiarone, A.~Giaretta, S.~T. Larsen, and
  S.~Dustdar.
\newblock Microservices: Migration of a mission critical system.
\newblock {\em IEEE Transactions on Services Computing}, pages 1--1, 2018.

\bibitem{mazzaraPhD}
Manuel Mazzara.
\newblock {\em Towards Abstractions for Web Services Composition}.
\newblock PhD thesis, University of Bologna, 2006.

\bibitem{Mazzara2010}
Manuel Mazzara.
\newblock Deriving specifications of dependable systems: toward a method.
\newblock {\em CoRR}, abs/1009.3911, 2010.

\bibitem{Mazzara11}
Manuel Mazzara, Faisal Abouzaid, Nicola Dragoni, and Anirban Bhattacharyya.
\newblock Toward design, modelling and analysis of dynamic workflow
  reconfigurations - {A} process algebra perspective.
\newblock In {\em Web Services and Formal Methods - 8th International Workshop,
  {WS-FM}}, pages 64--78, 2011.

\bibitem{MazzaraSDB2019}
Manuel Mazzara, Ilya Afanasyev, Smruti~R. Sarangi, Salvatore Distefano, and
  Vivek Kumar.
\newblock A reference architecture for smart and software-defined buildings.
\newblock {\em CoRR}, abs/1902.09464, 2019.

\bibitem{Mazzara2005}
Manuel Mazzara and Sergio Govoni.
\newblock {\em A Case Study of Web Services Orchestration}, pages 1--16.
\newblock Springer Berlin Heidelberg, 2005.

\bibitem{MazzaraSSE}
Manuel Mazzara, Kevin Khanda, Ruslan Mustafin, Victor Rivera, Larisa Safina,
  and Alberto Sillitti.
\newblock Microservices science and engineering.
\newblock In Paolo Ciancarini, Stanislav Litvinov, Angelo Messina, Alberto
  Sillitti, and Giancarlo Succi, editors, {\em Proceedings of 5th International
  Conference in Software Engineering for Defence Applications}, pages 11--20,
  Cham, 2018. Springer International Publishing.

\bibitem{MazzaraNSSU18}
Manuel Mazzara, Alexandr Naumchev, Larisa Safina, Alberto Sillitti, and
  Konstantin Urysov.
\newblock Teaching devops in corporate environments - an experience report.
\newblock In {\em Software Engineering Aspects of Continuous Development and
  New Paradigms of Software Production and Deployment - First International
  Workshop, {DEVOPS} 2018, Chateau de Villebrumier, France, March 5-6, 2018,
  Revised Selected Papers}, pages 100--111, 2018.

\bibitem{Meyer1988}
Bertrand Meyer.
\newblock {\em Object-Oriented Software Construction}.
\newblock Prentice-Hall, Inc., 1st edition, 1988.

\bibitem{roadmap}
Kizilov Mikhail, Antonio Bucchiarone, Manuel Mazzara, Larisa Safina, and
  V{\'{\i}}ctor Rivera.
\newblock Domain objects and microservices for systems development: a roadmap.
\newblock In {\em Proceedings of 5th International Conference in Software
  Engineering for Defence Applications}, 2017.

\bibitem{N15}
S.~Newman.
\newblock {\em Building microservices}.
\newblock O'Reilly Media, Inc., 2015.

\bibitem{nielsen}
Julie~Meinicke Nielsen.
\newblock {A Type System for the Jolie Language}.
\newblock Master's thesis, Technical University of Denmark, 2013.

\bibitem{predonzani2001}
{Predonzani P.}, {Sillitti A.}, and {Vernazza T.}
\newblock Components and data-flow applied to the integration of web services.
\newblock In {\em The 27th Annual Conference of the {IEEE} Industrial
  Electronics Society ({IECON}01)}, 2001.

\bibitem{Rice53}
Henry~Gordon Rice.
\newblock Classes of recursively enumerable sets and their decision problems.
\newblock {\em Trans. Amer. Math. Soc.}, 74:358--366, 1953.

\bibitem{rumbaugh1991object}
James Rumbaugh, Michael Blaha, William Premerlani, Frederick Eddy, William~E.
  Lorensen, et~al.
\newblock {\em Object-oriented modeling and design}, volume 199.
\newblock Prentice-hall Englewood Cliffs, NJ, 1991.

\bibitem{Salikhov2016b}
D.~Salikhov, K.~Khanda, K.~Gusmanov, M.~Mazzara, and N.~Mavridis.
\newblock Jolie good buildings: Internet of things for smart building
  infrastructure supporting concurrent apps utilizing distributed
  microservices.
\newblock In {\em Proceedings of the 1st International conference on Convergent
  Cognitive Information Technologies}, pages 48--53, 2016.

\bibitem{sillitti2002}
{Sillitti A.}, {Vernazza T.}, and {Succi G.}
\newblock Service oriented programming: a new paradigm of software reuse.
\newblock In {\em 7th International Conference on Software Reuse}, Lecture
  Notes in Computer Science 2319, pages 269--280. Springer Berlin Heidelberg,
  2002.

\bibitem{reftype}
Alexander Tchitchigin, Larisa Safina, Manuel Mazzara, Mohamed Elwakil, Fabrizio
  Montesi, and Victor Rivera.
\newblock Refinement types in jolie.
\newblock {\em Proceedings of the Institute for System Programming}, 28:33--44,
  2016.

\bibitem{YanCZM07}
Zhixian Yan, Emilia Cimpian, Michal Zaremba, and Manuel Mazzara.
\newblock {BPMO:} semantic business process modeling and {WSMO} extension.
\newblock In {\em 2007 {IEEE} International Conference on Web Services {(ICWS}
  2007), July 9-13, 2007, Salt Lake City, Utah, {USA}}, pages 1185--1186, 2007.

\bibitem{YanMCU07}
Zhixian Yan, Manuel Mazzara, Emilia Cimpian, and Alexander Urbanec.
\newblock Business process modeling: Classifications and perspectives.
\newblock In {\em Business Process and Services Computing: 1st International
  Working Conference on Business Process and Services Computing, {BPSC} 2007,
  September 25-26, 2007, Leipzig, Germany.}, page 222, 2007.

\end{thebibliography}
\end{document}